\documentclass[aps,twocolumn,prb]{revtex4-1}
\usepackage{graphicx,amssymb,amsmath}
\usepackage{graphicx}
\usepackage{dcolumn}
\usepackage{bm}
\usepackage[colorlinks,hyperindex, urlcolor=blue, linkcolor=blue,citecolor=black, linkbordercolor={.7 .8 .8}]{hyperref}

\newcommand{\bpar}{$B_{||}$}
\newcommand{\bperp}{$B_{\perp}$}
\newcommand{\btot}{$B_{tot}$}

\newcommand{\dl}{$d/\ell$}

\begin{document}
\title{Interlayer Interactions and the Fermi Energy of Bilayer Composite Fermion Metals}
\author{J. P. Eisenstein}
\affiliation{Institute for Quantum Information and Matter, Department of Physics, California Institute of Technology, Pasadena, CA 91125, USA}
\author{L. N. Pfeiffer}
\affiliation{Department of Electrical Engineering, Princeton University, Princeton, NJ 08544, USA}
\author{K. W. West}
\affiliation{Department of Electrical Engineering, Princeton University, Princeton, NJ 08544, USA}

\date{\today}

\begin{abstract}
When two 2D electron gas layers, each at Landau level filling factor $\nu=1/2$, are sufficiently close together a condensate of interlayer excitons emerges at low temperature.  Although the excitonic phase is qualitatively well understood, the incoherent phase just above the critical layer separation is not.  Using a combination of tunneling spectroscopy and conventional transport, we explore the incoherent phase in samples both near the phase boundary and further from it.  In the more closely spaced bilayers we find the electronic spectral functions narrower and the Fermi energy of the $\nu = 1/2$ composite fermion metal smaller than in the more widely separated bilayers.  We attribute these effects to a softening of the intralayer Coulomb interaction due to interlayer screening.

\end{abstract}
\maketitle

\section{Introduction}
At high magnetic field $B$ double layer two-dimensional electron systems (2DESs) can exhibit strongly correlated electronic phases which depend fundamentally on Coulomb interactions between electrons in opposite layers.  For example, in a bilayer 2DES in which the carrier density in each layer equals one-half the degeneracy of the lowest spin-resolved Landau level created by the magnetic field, the system will condense into an excitonic phase in which electrons in one layer are bound to holes in the other, provided that the layer separation and temperature are sufficiently small \cite{eisenstein2014}.  Conversely, if the separation between the layers is large, interlayer Coulomb interactions are weak and exciton condensation does not occur.  Nevertheless, Coulomb interactions between electrons in the same layer render the individual 2DESs very strongly correlated.  In the limit of very large layer separation each 2DES, in this half filling state, is well described as metallic phase \cite{halperin1993} of composite fermions (CFs) \cite{jain1989}, electrons to which two fictitious flux quanta are attached.   

Of interest here is the degree to which interlayer interactions at intermediate layer separations modify the compressible 2DES states in each layer.  This question is important since the precise nature of the phase transition to the excitonic phase remains poorly understood.  While this transition appears to be first-order (at least in some situations) \cite{schliemann2001,stern2002,zou2010}, the precise nature of the competing phase remains unclear.  

We report here a set of experiments, comprising interlayer tunneling spectroscopy and conventional magneto-transport, on two types of bilayer 2DES samples which differ dimensionally in only one way: the thickness of the barrier separating the two layers.  The samples with the narrower barrier allow for studies relatively close to the excitonic phase boundary, while the wider barrier samples provide access to the weakly coupled regime.  The direct comparison of tunneling and resisitivity data on these two classes of samples demonstrates that interlayer interactions (screening) soften the Coulomb repulsion between electrons within each layer.  This softening manifests as a narrowing of the electronic spectral functions of each layer, which are directly detected via the tunneling measurements, and a reduction in the Zeeman energy required to fully spin polarize the 2DES as observed in tilted field magneto-transport measurements.  Thus, while our measurements do not reveal qualitatively new 2DES phases at layer separations near the excitonic phase boundary, they do demonstrate that the energetics of each 2DES in the bilayer is significantly renormalized by interlayer interactions.  A simple model, based on dipolar interactions between electrons, is roughly consistent with the observed magnitude of this renormalization.

\section{Experimental}
The samples employed in this work are modulation-doped GaAs/AlGaAs heterostructures consisting of two GaAs quantum wells separated by a barrier layer of Al$_{x}$Ga$_{1-x}$As.  Two classes of such double quantum well (DQW) samples were grown and studied.  In one, the barrier separating the GaAs quantum wells is relatively narrow ($d_b=10$ nm) while in the other it is wide ($d_b=38$ nm) \cite{barrier}.   In both cases, the GaAs quantum wells are of width $w=18$ nm and are flanked by thick Al$_{0.32}$Ga$_{0.68}$As cladding layers.  Si delta-doping sheets are positioned in these cladding layers roughly 22 nm above and below the DQW.  These dopants populate the lowest subband in each quantum well with a 2DES of nominal density $n = 5.5 \times 10^{10}$ cm$^{-2}$.  As grown, the low temperature mobility of the 2DESs ranged from $\sim 1 \times 10^6$ cm$^2$/Vs in the $d_b=10$ nm samples to $\sim 2.5 \times 10^6$ cm$^2$/Vs in the $d_b=38$ nm samples.  The samples are patterned so that the 2DESs are confined to a 250 $\mu$m square region, with arms extending to ohmic contacts to the individual 2D layers \cite{eisenstein90}. These contacts enable both conventional magneto-transport measurements on the individual layers as well as direct measurements of the tunneling current $I$ flowing between the layers in response to an applied interlayer voltage $V$.  Independent control over the electron density in each layer is enabled by electrostatic gates on the top and back sides of the samples.  

For most of the data presented here, the 2D layer densities are tuned into equality \cite{density} and range from $n \approx 3.9$ to $7.3 \times 10^{10}$ cm$^{-2}$.   Over this range, the ratio of the center-to-center quantum well separation $d=d_b+w$ to the magnetic length $\ell=(\hbar/eB)^{1/2}$ at half-filling of the lowest Landau level is $2\lesssim d/\ell \lesssim 2.6$ for the narrow barrier sample and $3.9\lesssim d/\ell \lesssim 5.4$ for the wider barrier sample.  For comparison, the transition to the excitonic phase, observable in the narrow barrier samples at still lower densities \cite{eisenstein2014}, occurs near \dl\ =1.8.

\begin{figure}
\includegraphics[width=3 in]{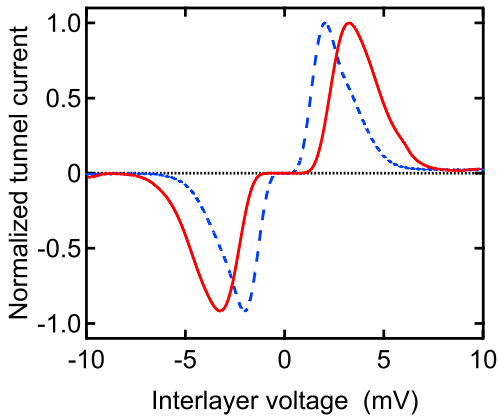}
\caption{\label{}(color online) Tunneling current-voltage characteristics at $\nu = 1/2$ (per layer) and $T=50$ mK in the narrow (blue, dashed) and wide (red, solid) barrier samples, at $B=4.13$ T and $B =4.24$ T, respectively.   The tunnel current has been normalized by its peak value at positive interlayer bias ($I_{peak} = 85$ pA and 1.04 nA for the narrow and wide barrier data, respectively.)}
\end{figure}

\section{Results}
Figure 1 displays typical interlayer tunneling current-voltage ($IV$) characteristics for both the narrow barrier (dashed blue trace) and the wide barrier samples (solid red trace) at high magnetic field and low temperature.  In both cases the Landau level filling fraction of the individual 2D layers is $\nu = nh/eB = 1/2$ (at zero interlayer bias \cite{density}). The applied magnetic field $B$ (and hence the per layer electron density) is very nearly the same in the two cases ($B=4.13$ $vs.$ 4.24 T).  Both traces exhibit well-known features of lowest Landau level interlayer tunneling:  A substantial suppression of the tunneling current around zero bias and a broad peak in the current at finite voltage \cite{eisenstein92,brown94,ashoori90}.  The suppression around zero bias is a Coulomb pseudogap arising from the inability of the interacting 2DES to rapidly accommodate the near-instantaneous injection (or withdrawal) of a tunneling electron at low energies, while the width of the peak at finite voltage reflects the interaction-driven broadening of the otherwise massively degenerate single-particle Landau level \cite{hatsugai93,he93,johannson93,efros93,kim94,varma94,haussmann96,levitov97,chowdhury2017,chowdhury2018}.  

In spite of these common features, interlayer tunneling in the wide and narrow barrier samples differs in ways both obvious and subtle. For example, as Fig. 1 makes clear, the pseudogap region of suppressed tunnel current around zero bias is broader, and the voltage location of the peak in the tunnel current is greater in the wide barrier sample than in the narrow barrier one.  Less obvious from the figure are systematic differences in the width of the tunneling peaks and in the nature of collapse of the tunnel current in the pseudogap region.   For our present purposes we focus on the width and voltage location of the tunneling peak.  

\begin{figure}
\includegraphics[width=3.1 in]{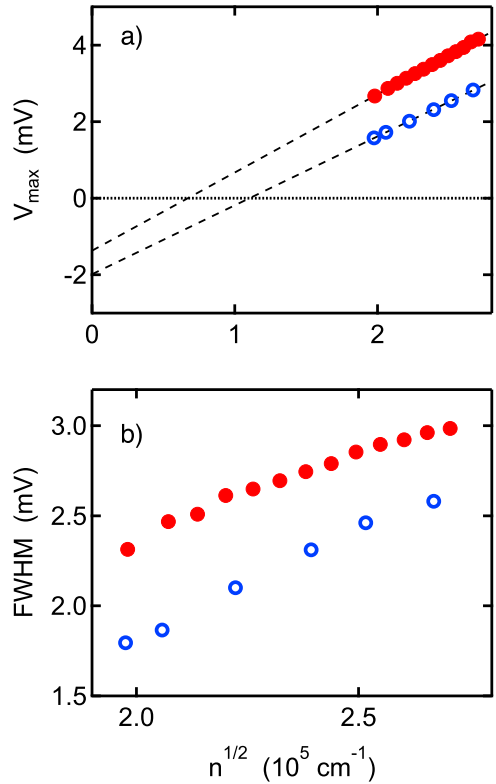}
\caption{\label{}(color online) a) Voltage location of the peak tunnel current $vs.$ $n^{1/2}$ in the narrow (open blue dots) and wide (solid red dots) barrier samples at $T = 50$ mK.  Dashed lines are linear least-squares fits, extrapolated to zero density.  b) Full width at half-maximum (FWHM) of tunneling peaks $vs.$ $n^{1/2}$.}
\end{figure}

Figure 2a displays the voltage location $V_{max}$ of the peak in the tunneling current at $\nu = 1/2$ versus the square root of the per layer electron density, $n^{1/2}$, for both the wide (red) and narrow (blue) barrier samples.  In both cases, the dependence is linear over the available data range, but extrapolates to a significant negative voltage, $V_{ex}$, in the zero density limit.  As reported and discussed previously, $V_{ex}$ is interpreted as arising from the final state excitonic attraction between a tunneled electron and the hole it leaves behind in the source 2D layer \cite{eisenstein95}.  In a simple model, one expects $V_{ex}=-\alpha e^2/\epsilon d$, with $\epsilon \approx 13 \epsilon_0$ the dielectric constant of the GaAs host and $\alpha$ a numerical factor dependent on the ratio \dl\ of the layer separation $d$ and the magnetic length $\ell$.  For the data in Fig. 2a, we find $\alpha \approx 0.5$ and $\alpha \approx 0.7$ for the narrow and wide barrier samples, respectively \cite{alpha}.  (As mentioned above, $d=d_b+w$, is the center-to-center separation between the quantum wells.)  That $\alpha$ is nearer to unity in the wider barrier samples makes sense since the charge defects become, in relative terms, more and more point-like as \dl\ is increases.

At $\nu = 1/2$ the mean intralayer Coulomb energy is of order $E_c=e^2/\epsilon \ell = (e^2/\epsilon) (4\pi n)^{1/2}$, ignoring small corrections arising from the finite thickness $w$ of the 2D layers and possible mixing with higher Landau levels.  Hence, if such interactions dominate the tunneling spectrum, it is not surprising that $V_{max}$ exhibits a linear dependence on $n^{1/2}$.  Interestingly, however, the different slopes of the data sets in Fig. 2a reveals that this scaling of $V_{max}$ with $n^{1/2}$ is sensitive to the separation $d$ between the quantum wells.  This is $not$ expected in a model of the tunneling process based upon independent 2D electron systems, modified only by a simple final state excitonic correction.  Writing $eV_{max}=eV_{ex}+\beta (e^2/\epsilon \ell)$, the fits to the data in Fig. 2a  reveal $\beta = 0.52$ and $\beta = 0.46$ for the wide and narrow barrier samples, respectively.  The coefficient $\beta$ reflects the strength of intralayer Coulomb interactions and its observed dependence on the layer separation $d$ indicates that those interactions are weaker in the narrow barrier sample than in the wider barrier one.  We attribute this weakening to enhanced screening arising from {\it interlayer} Coulomb interactions in the narrow barrier sample.  

Additional evidence for reduced intralayer Coulomb interactions in the narrow barrier tunnel junctions is illustrated in Fig. 2b, where the full width at half maximum (FWHM) of the tunneling peak is plotted $vs.$ $n^{1/2}$.  In the independent layer approximation, the tunneling peak represents a convolution of the Coulomb-broadened electronic spectral functions of the two 2D layers. That the tunneling peak widths in the narrow barrier sample are 15 - 25\% smaller than those in the wide barrier sample indicates a failure of this approximation and again suggests that interlayer screening softens the Coulomb repulsion between electrons in the same 2D layer.

To complement the preceding tunneling spectroscopic evidence that Coulomb interactions in a single 2DES are softened by the nearby presence of a second 2D layer, we turn to tilted field measurements of the ordinary longitudinal resistance $R_{xx}$.  It is well known that the spin polarization of a 2DES at $\nu = 1/2$ is incomplete at low electron density \cite{kukushkin99,dementyev99,melinte00,dujovne05,tracy07}.  Moreover, a transition from partial to complete spin polarization can be driven by adding an in-plane magnetic field \bpar\ to the perpendicular field \bperp\ which establishes $\nu = 1/2$.  The added in-plane magnetic field increases the spin Zeeman energy relative to the Coulomb energy $e^2/\epsilon \ell$ since the former depends on the total magnetic field \btot\ but the latter only on the fixed perpendicular field component, \bperp.  Ignoring additional effects of the in-plane field arising from the finite thickness of the 2D system \cite{thickness}, this increase of the Zeeman energy obviously favors maximal spin polarization.

In the Chern-Simons theory \cite{halperin1993} of the half-filled lowest Landau level, the strongly interacting electron system is approximated by a weakly interacting Fermi sea of composite fermions \cite{jain1989}.  At low electron density two such Fermi seas are present, one for ``up'' spin CFs and one for ``down'' spins.  The difference in the depths of these Fermi seas (i.e. their Fermi energies) is just the Zeeman energy $E_Z=|g| \mu_B B_{tot}$.  (Here $g\approx -0.44$ is the conduction band $g$-factor of GaAs and $\mu_B$ is the Bohr magneton.) As $E_Z$ is increased, via tilting at fixed \bperp, relative to the Fermi energies, the minority spin band depopulates and the 2DES becomes fully polarized.  This transition occurs when the Zeeman energy $E_Z$ matches the Fermi energy $E_F$ of the majority spin CFs \cite{park98,zhang2017}.  In a clean 2DES the Fermi energy of CFs at $\nu = 1/2$ is determined entirely by Coulomb interactions: $E_F=\gamma e^2/\epsilon \ell$, with $\gamma$ a numerical factor of order unity \cite{gamma}.  Conveniently, experiments have shown that $R_{xx}$ at $\nu = 1/2$ increases steadily as \bpar\ is applied, but then saturates when the spin polarization is complete \cite{li09}.  Hence, the total magnetic field $B_{tot}^*$ at which saturation sets in provides a transport determination of the CF Fermi energy: $E_F =|g|\mu_B B_{tot}^*$. 

\begin{figure}
\includegraphics[width=3.1 in]{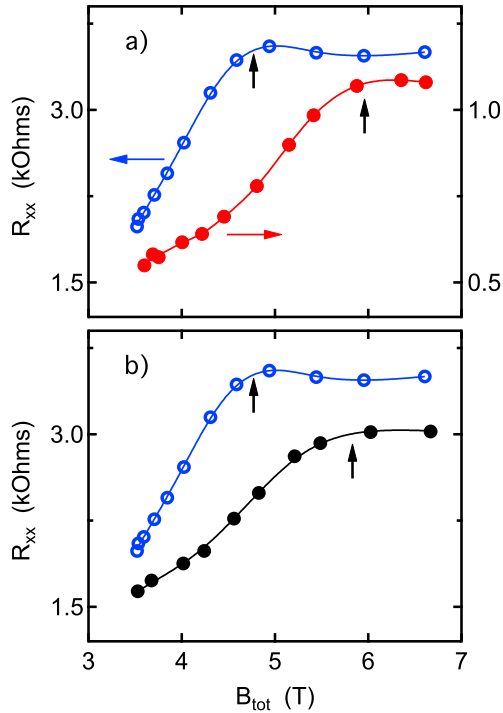}
\caption{\label{}(color online) Tilted field response of the longitudinal resistance $R_{xx}$ at $\nu = 1/2$ and $T=50$ mK. Perpendicular magnetic field fixed at $B_{\perp}\approx 3.56$ T.  a) Comparison between narrow barrier (open blue dots) and wide barrier (solid red dots) samples.  Both 2D layers at $\nu = 1/2$ with the resistance measured in one of the layers.  b) Comparison of narrow barrier sample with both layers at $\nu =1/2$ (open blue dots) $vs.$ situation with one layer at $\nu = 1/2$ and the other fully depleted (solid black dots).  Upward arrows suggest transition points, $B_{tot}^*$, to full spin polarization.}
\end{figure}

Figure 3a compares tilted field measurements of $R_{xx}$ at $\nu = 1/2$ for the narrow and wide barrier samples.  The samples are density balanced, $\nu =1/2$ in both 2D layers, but $R_{xx}$ is measured with current (typically 1 nA) flowing in only one of the two layers.  In order to fairly compare the samples, their carrier densities were adjusted to near equality:  $n = 4.25$ $vs.$ $4.35 \times 10^{10}$ cm$^{-2}$, per layer, for the narrow and wide barrier samples, respectively.   Both samples show $R_{xx}$ rising steadily with \btot\ (with \bperp\ fixed) before saturating at a resistance roughly twice that observed at \bpar=0.  Interestingly, the ``knee'' in the resistance occurs near $B_{tot}^* \approx 4.8$ T in the narrow barrier data but at about  $B_{tot}^* \approx 6$ T in the wide barrier case. This implies that the CF Fermi energy in the narrow barrier sample is roughly 20\% smaller than in the wide barrier sample.

As a check on the above conclusions, the top and backside gates on the narrow barrier sample were adjusted so that only one of its two quantum wells contained a 2DES and the density of that 2DES was set to the same value ($n=4.25 \times 10^{10}$ cm$^{-2}$) as in the density balanced situation just discussed.  Once again, the tilted field dependence of $R_{xx}$ at $\nu = 1/2$ was measured.  As Fig. 3b demonstrates, this arrangement led to essentially the same total magnetic field $B_{tot}^*$ needed to fully polarize the electron spins in the 2DES as found in the wide barrier, density balanced, bilayer sample.  Moreover, the general shape of the $R_{xx}$ $vs.$ \btot\ dependence more closely resembles that found in the wide barrier sample than in the same narrow barrier sample with both layers at $\nu =1/2$.  These observations strongly support our conclusion that the different spin polarization fields $B_{tot}^*$ found in the narrow and wide barrier bilayer samples is a genuine interlayer interaction effect, and not an artifact arising from the comparison of distinct heterostructure samples.  Finally, these results indicate that the effectiveness of interlayer screening attenuates quickly with increasing layer separation.

\section{Summary and Conclusions}
The tunneling spectroscopy and magneto-resistance measurements described here are mutually consistent and support our conclusion that interlayer screening substantially softens intralayer Coulomb interactions and reduces the CF Fermi energy in closely spaced bilayer 2D systems.  The magnitude and character of such softening is determined both by the distance between the layers and the physical properties (compressibility, conductivity, etc.) of the screening layer.  In the present instance, with both layers at $\nu = 1/2$, each 2DES is a compressible, conducting quantum fluid. Hence, interlayer screening at some level should be present.   If, in considering the electron-electron interactions in one of the layers, the other is simply treated as a perfectly conducting plane, then the elementary concept of image charges suggests that those interactions become dipole-like, thus strongly suppressing the long range coulombic repulsion between electrons.  In this highly over-simplified model the magnitude of this suppression is quite substantial. For example, the repulsive force between two point-like electrons separated by $r=n^{-1/2}=2\sqrt{\pi}\ell$ (at $\nu=1/2$) is reduced by almost 30\% if a perfectly conducting parallel metallic plane is positioned a distance $d=2\ell$ away.  Of course, the 2DES at $\nu = 1/2$, while compressible, is not a perfect metal and the resulting screening may be less.  Our tunneling and tilted field resistivity data suggest that in the narrow barrier sample, where $2 \lesssim d/\ell \lesssim 2.6$, the mean intralayer Coulomb energy is suppressed by 15-30\%, relative to its value in the wide barrier sample, where $3.9 \lesssim d/\ell \lesssim 5.4$.   This is roughly consistent with the naive dipolar model mentioned above.  

In conclusion, tunneling spectroscopy and magnetoresistance measurements provide quantitative evidence that interlayer Coulomb interactions can effectively screen intralayer interactions in closely spaced bilayer 2D electron systems. While we find that down to effective layer separations $d/\ell \approx 2$, the bilayer system at total filling factor $\nu_T=1/2+1/2=1$ appears to remain well described as two parallel composite fermion metals, the energetic parameters (e.g. the Fermi energy) of these metallic states are significantly renormalized by interlayer Coulomb interactions.  

\begin{acknowledgements}
It is a pleasure to acknowledge helpful discussions with Gil Refael and Jainendra Jain. This work was supported in part by the Institute for Quantum Information and Matter, an NSF Physics Frontiers Center with support of the Gordon and Betty Moore Foundation through Grant No. GBMF1250.  The work at Princeton University was funded by the Gordon and Betty Moore Foundation through Grant GBMF 4420, and by the National Science Foundation MRSEC Grant 1420541.
\end{acknowledgements}


\begin{references}
\bibitem{barrier}  In order that the tunneling rate between the quantum wells be comparable, the tunnel barrier height was adjusted via the Al concentration $x$.  For the 10 nm barrier samples $x=0.90$, while for the 38 nm barriers $x=0.10$.
\bibitem{density} The equal density condition is detected via the voltage symmetry of the interlayer tunnel resonance at zero magnetic field. Finite interlayer bias disrupts this density balance but the effect is small and is ignored here.
\bibitem{thickness} For the spin polarization transition at $\nu = 1/2$ such finite thickness effects have been carefully studied \cite{li09} and found to be of secondary importance.
\end{references}
\end{document}